\documentclass[epj]{webofc}
\usepackage[utf8]{inputenc}
\usepackage[varg]{txfonts}   
\usepackage{booktabs}
\usepackage{xcolor}
\definecolor{darkred}{rgb}{0.4,0.0,0.0}
\definecolor{darkgreen}{rgb}{0.0,0.4,0.0}
\definecolor{darkblue}{rgb}{0.0,0.0,0.4}
\usepackage[bookmarks,linktocpage,colorlinks,
    linkcolor = darkred,
    urlcolor  = darkblue,
    citecolor = darkgreen]{hyperref}
\usepackage{subfigure}
\wocname{EPJ Web of Conferences}
\woctitle{Lattice2017}
\usepackage{listings}

\begin{document}
%
\selectlanguage{english}
\title{%
Wilson and Domainwall Kernels on Oakforest-PACS
}
\author{%
\firstname{Issaku} \lastname{Kanamori}\inst{1}\fnsep\thanks{Speaker, \email{kanamori@hiroshima-u.ac.jp}%
} \and
\firstname{Hideo} \lastname{Matsufuru}\inst{2}
}
\institute{%
Department of Physics,
Hiroshima University,
Higashi-hiroshima 739-8526, Japan
\and
Computing Research Center,
High Energy Accelerator Research Organization (KEK),
Oho 1-1, Tsukuba 305-0801, Japan
}
\abstract{%

We report the performance of Wilson and Domainwall Kernels on
a new Intel Xeon Phi Knights Landing based machine named Oakforest-PACS,
which is co-hosted by University of Tokyo and Tsukuba University and
is currently fastest in Japan.
This machine uses Intel Omni-Path for the internode network.
We compare performance with several types of implementation including
that makes use of the Grid library.
The code is incorporated with the code set Bridge++.
}
\maketitle
\section{Introduction}\label{intro}

It is crucially important for the lattice QCD to have
simulation codes which run efficiently on new machines.
Since we have a plenty of codes developed so far,
it is desirable to have a practical performance with only small changes
of the existing codes.
Our main goals of this work are to understand the behavior of the
machine, (minimal) recipe for the better performance, and to examine
portable code design.
Rather than winning the performance competition on the machine,
we are aiming at reducing the tuning cost for reasonably high
performance on new machines.

A trend of the modern high performance computers is to have more and
more cores in a single processor.
An extreme case is GPUs, which has $O(1000)$ cores and 
works as an accelerator attached to a host CPU.
The Intel Xeon Phi series is another example:
the number of cores is more than that of the standard CPUs like Intel
Xeon processors while much less than that of GPUs.
Contrary to the first generation of Xeon Phi named Knights Corner,
the second generation, Knights Landing (KNL), no longer requires
a host CPU and works as a many core CPU.

In this work, we explore performance tuning on a KNL-based cluster
machine, the Oakforest-PACS system.
A typical performance bottleneck of lattice QCD simulations
is multiplication of a Dirac operator in a linear equation solver
to determine the fermion propagator.
We therefore start with tuning of typical Dirac operators,
the Wilson and domainwall fermions.
We examine two independent implementations:
one is rather simple (denoted by impl-1), while the other is more
aggressive (impl-2).
These implementations are done by extending an existing lattice QCD
code set, Bridge++ \cite{bridge}.
The performance is measured on the Oakforest-PACS system so as to
examine the strong and weak scaling.
For the other recent code developments, see the plenary talk
by Rago~\cite{thiscontrib_plenary} and references therein.

This paper is organized as follows.
The next section briefly explains our target machine.
In Sect.~\ref{sec:implementation}, we describe our implementations
in detail.
The benchmark result is presented in Sect.~\ref{sec:benchmark}.
Finally, Sect.~\ref{sec:conclusions} gives our concluding remarks.

\section{Target Machine}
\label{sec:machine}

Our target machine is the Oakforest-PACS system hosted by the Joint
Center for Advanced High Performance Computing
(JCAHPC, University of Tokyo and University of Tsukuba, Japan)
\cite{Oakforest-PACS_website}.
It started full operation in April 2017 and ranked as the 7th
(1st in Japan) in the Top 500 in June 2017.
The system is made of 8208 nodes each having one Intel Xeon Phi 7250
(68 cores, 1.4GHz, KNL architecture) processor.
The internode connection is the Intel Omni-Path with the full-bisection
fat tree structure.

The KNL architecture has SIMD vector registers of 512 bits length,
which can process 16 single or 8 double precision floating point numbers
simultaneously with the AVX-512 instruction set.
One of the key features of the KNL is the memory structure:
it has 16GB MCDRAM in between the cache and main memory.
Its bandwidth of more than 400 GB/s is much larger than the DDR4
memory of 90 GB/s.
The MCDRAM can work either as a cache, a part of main memory, or their
hybrid.
The L2 cache is shared by two cores that constitutes a unit called tile.
For more details on the KNL architecture, see for example
Refs.~\cite{KNLtextbook, thiscontrib_intel}.

\section{Implementation Details}
\label{sec:implementation}

It is not realistic to develop a new code on each new machine from
scratch.
As a framework of our development, we use the Bridge++ code set
\cite{bridge}.
It is described in C++ with object oriented manner intending to be
readable, extendable, portable and with practically high performance.
Since the original Bridge++ code is implemented with a fixed field
data layout and only in double precision, we extend the code by
employing the C++ template so as to enable arbitrary index ordering of
the field and any type of the real number.
This is also convenient for the tuning specific to an architecture.
Thanks to C++ template techniques, while general code is kept
working, machine specific tuning is incorporated by specialization
of the template.
Such extension is applied to the most time consuming part,
{\it e.g.} the linear equation solver, while keeping the framework
and measurement codes of Bridge++ available.
This strategy was previously adopted to use GPUs with OpenCL and
OpenACC \cite{Motoki:2016rii, bridge_alt}.
We apply this extended-Bridge++ approach to KNL with modifications
proper to the many-core and SIMD architecture.

\begin{figure}[thb]
  \centering
 \raisebox{5mm}{%
 \includegraphics[scale=0.4]{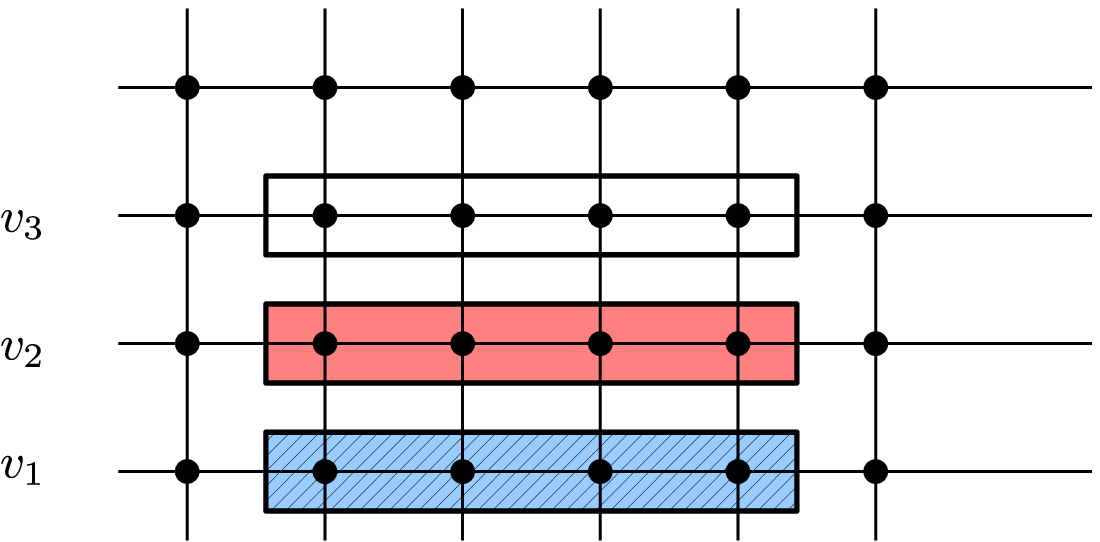}%
}
 \hfil
  \includegraphics[scale=0.4]{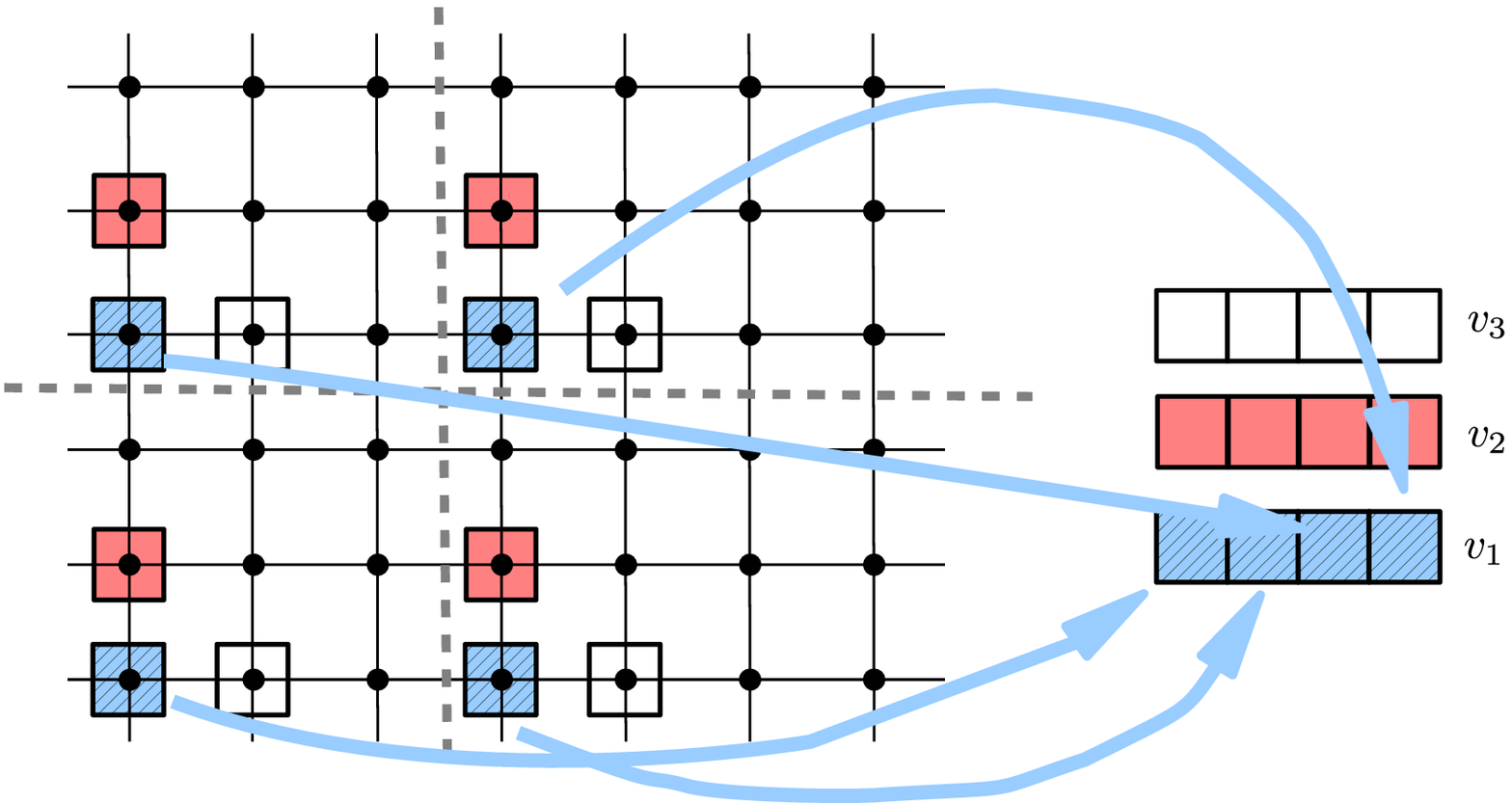}
 \caption{Packing data into SIMD variables for double precision.
 The left panel is for the
 impl-1 (1-dim. packing) and the right panel for the impl-2
 (Grid-like packing).
 For single precision, the length for the packing becomes twice
 longer (impl-1) or the subdomain division becomes 3-dimensional one (impl-2).}
  \label{fig:simd_lattice}
\end{figure}

To fully make use of the KNL architecture, one needs to change
the data layout so as to match the SIMD vector of 512-bit length
and apply the AVX-512 intrinsics to the code.
We store complex numbers as an array-of-structure, which is
compatible to an array of \texttt{std::complex<double>}
or \texttt{std::complex<float>}, so that one SIMD vector
handles 4 double or 8 single precision complex numbers.
There is still a variety of packing the lattice field degrees of
freedom into one SIMD vector.
To investigate how the details of the implementation affect the
performance, we prepare the following two implementations.
Their main differences lie in the data layout and how the Intel
AVX-512 intrinsics are applied.
\begin{description}

\item[Impl-1 (simple):]
The first implementation intends to be as simple as possible and
avoid using additional libraries.
As a natural modification of the original Bridge++ data layout,
the data on the continuous sites in $x$-direction are packed in a
single SIMD vector, as shown in the left panel of
figure~\ref{fig:simd_lattice}.
The hopping to the $x$-direction requires a special care in
rearranging the data inside a SIMD vector.
In addition, the number of lattice sites in this direction must be
a multiple of $4$ (in double precision) or 8 (single) on each node.
For simplicity, we do not parallelize this code in $x$-direction.
In this implementation, the AVX-512 intrinsics is manually applied
just by wrapping macros and inline functions. 

\item[Impl-2 (aggressive):]
The second implementation employs more advanced tuning techniques
than the first one.
The impl-2 adopts a data structure similar to the Grid library
\cite{Boyle:2016lbp, grid}.
The local lattice in a node is divided into $2\times 2$
(in double precision) or $2\times 2\times 2$ (single) subdomains.
From each subdomain, data on a site is packed into a SIMD vector,
as sketched in the right panel of figure~\ref{fig:simd_lattice}.
As for the application of the AVX-512 intrinsics, we make use of
an existing library.
We have found that \texttt{simd} directory in Grid
\cite{Boyle:2016lbp, grid}
provides an appropriate wrapper of the complex arithmetics.
In figure~\ref{fig:simd_codes} we list example codes to execute
$c_i = a_i^*b_i$ with $i=1,\dots,4$ simultaneously.

\end{description}

\begin{figure}[thb]
  \centering
{\small
 \begin{lstlisting}[frame=single, language=C++]
 // a, b are vector variables, i.e., __m512d
  __m512d a_real = _mm512_shuffle_pd( a, a, 0x00 );
  __m512d a_imag = _mm512_shuffle_pd( a, a, 0xFF );
  a_imag = _mm512_mul_pd( a_imag, _mm512_permute_pd( b, 0x55 ) );
  __m512d c =_mm512_fmsubadd_pd( a_real, b, a_imag );      

  // a, b are vComplexD defined in simd directory in Grid
  vComplexD c=conj(a)*b;
 \end{lstlisting}
}
 \caption{Example codes for $c_i=a_i^* b_i$ ($i=1,\dots, 4$), where complex numbers
 $a_i$, $b_i$, and $c_i$
  are stored in \texttt{a}, \texttt{b} and \texttt{c},
 respectively.
 The upper implementation is with intrinsics, and the lower one uses
 vector variables provided by Grid.}
 \label{fig:simd_codes}
\end{figure}

In both the implementations, the parallelization is done
in accord with {\tt MPI\_THREAD\_FUNNELED}, {\it i.e.} only the master
thread performs the communication.
The ways to assign the tasks to OpenMP threads are different in
the impl-1 and impl-2.
The master thread takes also charge of computation in impl-1,
while concentrates on the communication in impl-2
(except for the 1 thread/process case).
The modification is simple and its effect will be estimated
separately in near future.

In impl-2, we further employ the following two tuning techniques
and test for the Wilson fermion.
The first one is manual prefetching.
Inserting prefetch command (and disabling automatic prefetch by
compiler), prefetching data to L1 and L2 cache can be controlled
manually.
The other technique is loop tiling, which is implemented as follows.
In impl-2 we separate the `boundary' from the `bulk' only for the
contribution in the
$z$- and $t$-directions, while the loop tiling is done in $x$- and
$y$-directions which are the inner coordinates in our lattice site
index ordering.
During the communication in $x$- and $y$-directions, data packing
for the communication in $z$- and $t$-directions is overlapped.
The bulk computation overlaps with the communication in $z$- and
$t$-directions.
In this way, we implement the loop tiling independently of
the bulk/boundary separation, as making implementation easier.
Table~\ref{tab:impls} summaries the implementations for both
impl-1 and 2.

For the domainwall fermion operator, there are two possible
implementations.
One is to treat the 5th direction as an external degrees of freedom
and to apply the Wilson operator repeatedly.
The other is to treat the 5th direction as an internal degrees of
freedom on each site by describing a dedicated code for the
domainwall operator.
In the extended Bridge++ code, the former is already available
as a template class that is commonly applicable to the GPU code.
This is applicable to both the impl-1 and 2. 
The latter approach would achieve better performance owing to
more efficient reuse of the gauge field in the cache.
The latter code is developed for impl-2, while optimization
is still at a preliminary level.

\begin{table}[htb]
\caption{Summary of implementation 1 (simple) and 2 (aggressive).}
\label{tab:impls}
 \centering
 \begin{tabular}{lll}
\toprule
implementation  &  1 (simple) & 2 (aggressive) \\
\midrule
  arithmetics of SIMD & low level intrinsics  & library (Grid) \\
  data packing  & 1 dim. continuous &  multi dim. a la Grid \\
  maximum MPI parallelizing & 3 dim. & 4 dim. \\
  manual prefetch      & no & yes (no for domainwall)   \\
  loop tiling   & no & 2-dim tiling (no for domainwall)   \\
  domainwall    & using Wilson & native implementation \\ 
  \bottomrule
 \end{tabular}
\end{table}

\section{Benchmark Results}
\label{sec:benchmark}

The benchmark teats are performed with the following setting.
Among 68 cores in a single node, we use only 64 cores for computation
and leave remaining 4 cores to OS.
The affinity for the MCDRAM is cache quadrant.
The performance is determined by measuring elapsed time for
1,000 times multiplications of
the Dirac operator.
The size of the 5th direction for the domainwall operator is set to 8.

\subsection{Wilson operator}

Let us start with the Wilson Dirac operator.
The performance on a single node without MPI communication (compiled
without MPI) on $32^3\times 64$ lattice results in
241 GFlops (single precision, 4 threads/core: totally 256 threads)
and
147 GFlops (double, 4 threads/core)
for the impl-1, and
339 GFlops (single, 2 threads/core)
and
174 GFlops (double, 2 threads/core)
for the impl-2.
We observe clear dependence on the number of the
threads/core for impl-1, which indicates that having the more
threads/core is the faster: for 1 thread/core, only 86 GFlops is achieved
in single precision.
On the other hand, the impl-2 has mild dependence on the number of
threads per core, that amounts roughly up to 10\% of difference,
and in most cases 1 or 2 threads/core is the fastest.
In the following, we adopt the 4 thread/core case for impl-1
and pick up the fastest case from 1-4 threads/core for impl-2.

We first examine the effect of the manual prefetch.
We observe that the manual prefetch is efficient for a single node
job and increases the performance more than 20\%.
As the number of nodes increases, however, the effects becomes smaller
and only a few \% at 16 nodes. 
Since our supposed size of typical simulations is of $O(10)$ nodes
or more, it seems not worth applying the manual prefetching.

\begin{figure}[thb]
  \centering
 \includegraphics[width=0.45\linewidth]{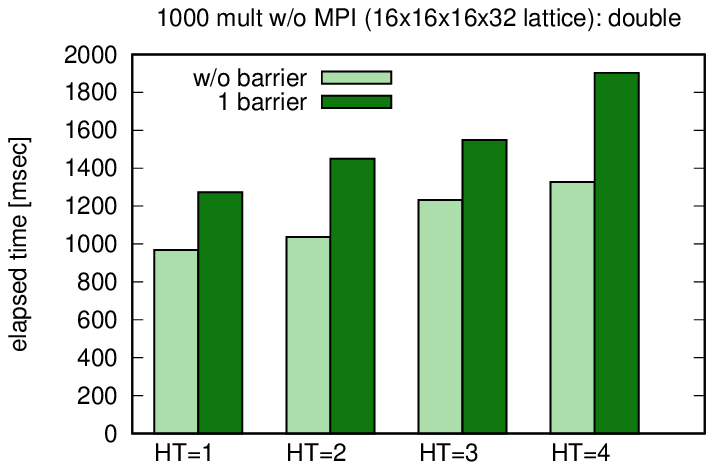}
 \quad
 \includegraphics[width=0.45\linewidth]{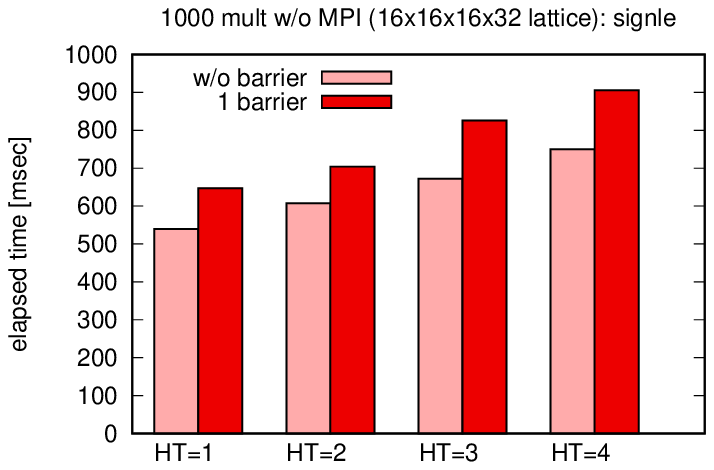}
 \caption{
Elapsed time for 1000 multiplications of the Wilson Dirac operators with
various number of threads/core labeled by HT (hyperthreading)
in the double (left panel) and single (right) precision.
The dark and light colored bars are results with and without barrier
synchronization.}
 \label{fig:timing_barrier}
\end{figure}

It is useful to quantify the effect of synchronization of threads
on the performance.
Figure~\ref{fig:timing_barrier} shows impact of single insertion
of \texttt{\#pragma omp barrier} in the Wilson multiplication
on a single node without MPI communications for the impl-2.
The elapsed time receives sizable effect by the barrier synchronization.
Note that the correct result is obtained by inserting synchronizations.
One barrier seems to cost 0.1--0.5 msec of penalty, which is far from
negligible\footnote{
After removing the loop tiling, however, such a big difference
between with and without barriers disappears.
This strongly suggests that the synchronization costs was due to a load
imbalance caused by the loop tiling (
I.K. thanks H. Servat from Intel for his suggestion during the conference
to investigate the load imbalance).
In our implementation, the loop outside the tiling is thread parallelized,
and a large chunk size can easily cause a load imbalance if the number
of threads is not a multiple of number of the chunks for tiling.}.

\begin{figure}[thb]
 \centering
 \sidecaption
 \includegraphics[width=0.55\linewidth]{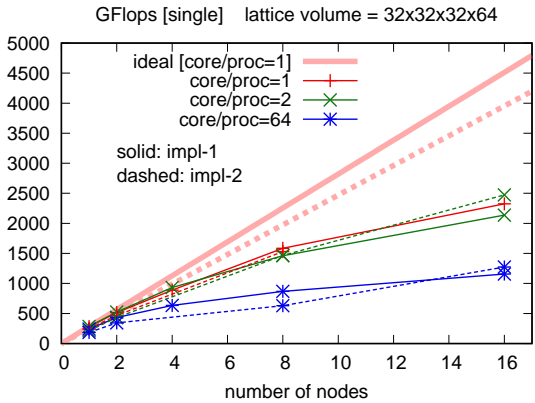}
 \caption{
 Strong scaling plot of the Wilson operator multiplication on
 a $32^3\times 64$ lattice.
 The solid and dashed lines represent the results of the impl-1 and
 impl-2, respectively.
 As a reference, ideal scalings from single node results are plotted
 with pink thick lines.}
 \label{fig:strong_scaling_wilson}
\end{figure}

\begin{figure}[thb]
 \centering
 \includegraphics[width=0.5\linewidth]{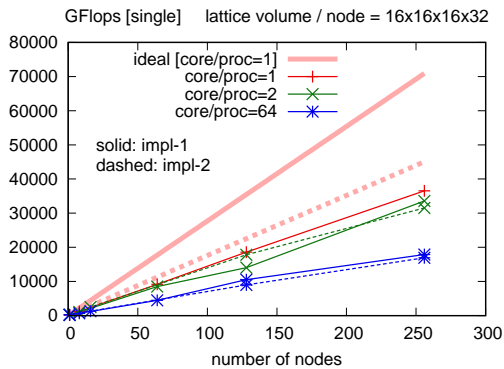}
\hspace{-0.02\linewidth}
 \includegraphics[width=0.5\linewidth]{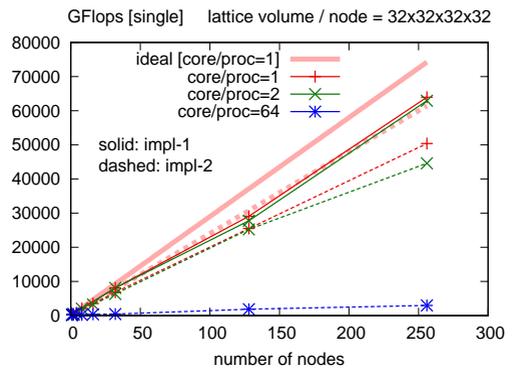}
 \caption{
 Weak scaling plot of the Wilson operator multiplication on
 $16^3\times 32$ (left panel) and $32^3\times 32$ (right) lattices.
 The solid and dashed lines represent the results of the impl-1 and
 impl-2, respectively.
 Ideal scalings from the single node results are plotted
 with pink thick lines.}
 \label{fig:weak_scaling_wilson}
\end{figure}

We now turn on the MPI communication and  observe the scaling
behavior against the number of KNL nodes.
Figures~\ref{fig:strong_scaling_wilson} and \ref{fig:weak_scaling_wilson}
respectively show the strong and weak scaling plots
of the multiplication of the Wilson Dirac operator.
We measure the performance with 3 different settings with respect to
the number of cores per MPI process, 1, 2 and 64:
The 1 core/proc.\ (red \texttt{$+$} symbol) case is equivalent to
64 MPI processes on each node and corresponds to the flat MPI.
The 2 cores/proc.\ (green \texttt{$\times$}) case assigns 1 MPI
process to a tile.
The 64 cores/proc.\ (blue \texttt{$\ast$}) case has 1 MPI process
on each node.
In the figure, we also plot the ideal scaling from the single node
result.
Note that since the scaling plot are measured with the program
generally including the communication, which implies the copy of
the boundary data is performed independently of the number of nodes.
Thus the single node performance of the scaling plot is different
from the performance without MPI stated in the beginning of this
section, due to redundant treatment of the boundaries.

In spite of its simplicity, the performance of impl-1 (solid line) is
comparable to the performance of impl-2 (dashed line) in most cases.
In both the implementations, the 1 and 2 cores/proc.\ cases tend to
exhibit better performance than the 64 cores/proc.\ case.
In the strong scaling plot, figure~\ref{fig:strong_scaling_wilson},
the performance of the former two cases on 16 nodes reduces to
roughly half the ideal scaling suggested by the single node result.

In the weak scaling plots in figure~\ref{fig:weak_scaling_wilson},
clear dependence on the lattice size in each node is observed.
For smaller lattice ($16^3 \times 32$ lattice/node), the slopes of
of performance of 1 and 2 cores/proc.\ are almost halved.
For larger lattice ($32^4$ lattice/node), the results of 1 and 2
cores/proc.\ show better scaling behavior.
The impl-1 scales with more than 80\% of the ideal scaling
up to 256 nodes at which it results in about 240 GFlops/node.
On the other hand, the 64 cores/process case does not scale at all
for $32^4$ lattice/node, in which we abandon to measure
the performance with impl-1.
A large fluctuation of the elapsed time for the 64 core/proc.\ case
is also observed.
We have not understood what causes these behaviors.

\subsection{Domainwall operator}

For the domainwall operator multiplication,
the optimal numbers of threads per core on a single node
are evaluated on a $32^3\times 64$ lattice as follows.
For the impl-1, the domainwall operator is composed of the Wilson
operator and linear algebraic functions.
The best performances is obtained with 64 MPI processes/node
and 4 threads/core and result in 89 GFlops (single precision)
and 58 GFlops (double).
In the case of impl-2, the 5th direction is integrated into
the on-site degrees of freedom.
The best performance is obtained with 32 MPI processes/node and
4 threads/core as
395 GFlops (single precision) and 197 GFlops (double).
As expected, the native implementation to domainwall (impl-2)
is much faster.
In the following, we concentrate on impl-2.

\begin{figure}[thb]
 \centering
 \includegraphics[width=0.5\linewidth]{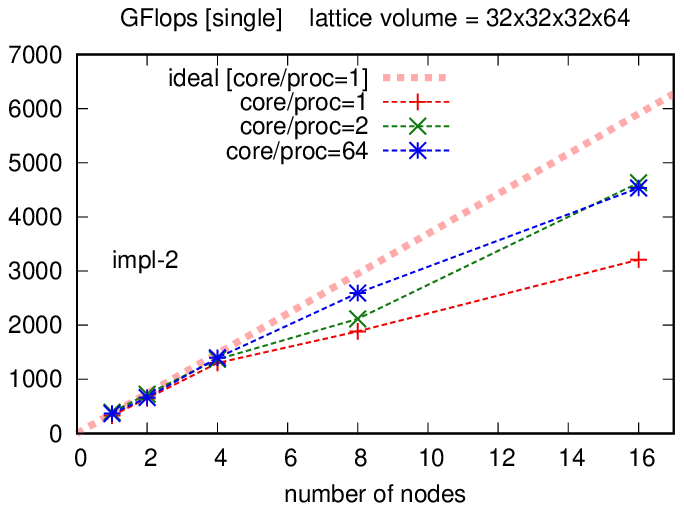}
\hspace{-0.02\linewidth}
 \includegraphics[width=0.5\linewidth]{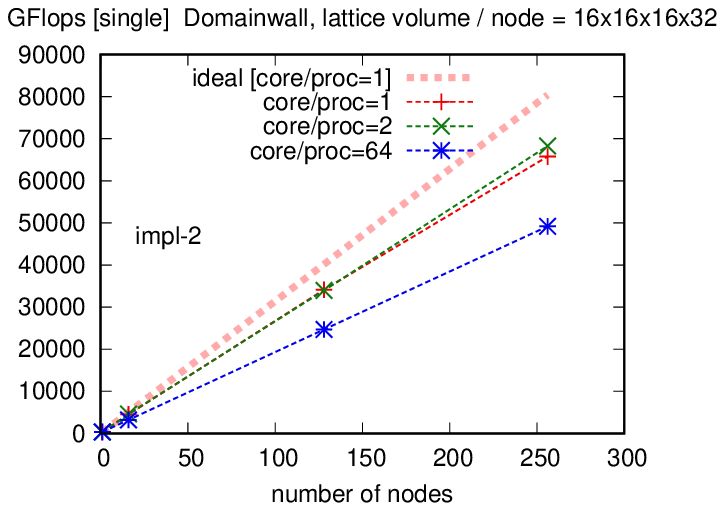}
 \caption{
 Strong scaling on $32^3 \times 64$ lattice (left panel)
 and weak scaling on $16^3\times 32$ lattice/node (right panel)
 of the Domainwall  multiplication.
 The results of impl-2 are plotted.
 As a reference, ideal scalings from single node results are plotted
 with pink thick lines. }
 \label{fig:scaling_domainwall}
\end{figure}

Figure~\ref{fig:scaling_domainwall}
shows the strong and weak scalings measured in a similar manner as
the Wilson case.
The reference performance of the ideal scaling is plotted using
the result with 1 core per MPI process on a single node:
369 GFlops/node on $32^3\times 64$ and 314 GFlops/node on
$16^3 \times 32$ lattices.
In the weak scaling plot, good scaling is observed up to 256 nodes
with more than 250 GFlops/node.
Compared to the Wilson case, the performance with 64 cores/process
({\it i.e.} 1 MPI process per node) is not much behind the others.
We observe, however, a fluctuation of the performance of more than
20 times over different number of threads/core\footnote{
In addition, with slightly modified code, a similar fluctuation
is observed in run by run \cite{cander2017}.}.
The origin of such fluctuations are now under investigation.

\section{Conclusions and Discussions}
\label{sec:conclusions}

We showed our benchmark results of the Wilson and domainwall
operator multiplication on a parallel KNL cluster, Oakforest-PACS.
Based on the extended Bridge++ code,
we develop two implementations: simple one with proper code
(impl-1) and aggressive one that exploits a part of Grid library
for the SIMD operations (impl-2).
In both the implementations, the codes are incorporated into
the Bridge++ framework so as to be called with common interface.

For the Wilson operator multiplication we investigated the effects
of implementation in some detail, and found that impl-1 and
impl-2 give similar performance.
This result indicates that the key ingredients of the tuning for KNL
are the data layout suitable to the SIMD variables, and
the use the AVX-512 intrinsics.
Code design that enables to easily incorporate such optimization
procedures is helpful to extend the tuning to other observables.
Our another finding is about manual prefetching.
It was efficient only on a single or a few nodes runs.
Unless we need optimized performance specialized to a single
node run, it is not worth spending time for trial and error for
the prefetch tuning.

For the domainwall operator, we first found that for better
performance the 5th dimensional degree of freedom should be
integrated into the on-site degrees of freedom, as implemented
in impl-2.
The scaling behavior is observed for impl-2.
As a common tendency to the Wilson and domainwall operators,
the 1 and 2 cores/process cases give better performance than
the 64 cores/process case.
This implies that the KNL cluster can be efficiently used as
a massively parallel machine.

Although the performance is not very optimal, it is reasonable
compared to the cost of implementation.
Note that considering that the flop-per-byte of the Wilson operator
is 1.12 Flop/Byte for single precision and 400 GB/s of MCDRAM
bandwidth, we expect 357 GFlops/node, while the performance
may becomes better by cache reusing.
Our sustained performance on a single node is only slightly
bellow this value.
Our results including the performance of the iterative solvers
will appear in \cite{cander2017}.

\subsection*{Acknowledgments}

We thank Peter Boyle, Guido Cossu,
Ken-Ichi Ishikawa, Daniel Richtmann, Tilo Wettig,
and the members of Bridge++ project for valuable discussion.
The numerical simulations were performed on Oakforest-PACS system
hosted by JCAHPC, with support of Interdisciplinary Computational Science Program in CCS, University of Tsukuba.
This work is supported by Priority Issue 9 to be
tackled by Using Post K Computer, by Joint Institute for
Computational Fundamental Science (JICFuS),
and by JSPS KAKENHI (Grant Numbers JP25400284, JP16H03988), 

\bibliography{lattice2017}

\begin{thebibliography}{10}

\bibitem{bridge}
{Bridge++ project}, \urlstyle{tt}\url{{http://bridge.kek.jp/Lattice-code/}}

\bibitem{thiscontrib_plenary}
A.~Rago, \emph{{Lattice QCD on new chips: a community summary}}, in
  \emph{Proceedings, \href{http://inspirehep.net/record/1425631}{35th
  International Symposium on Lattice Field Theory (Lattice2017)}: Granada,
  Spain}, to appear in EPJ Web Conf.

\bibitem{Oakforest-PACS_website}
{Joint Center for Advanced High Performance Computing (JCAHPC)},
  \urlstyle{tt}\url{{https://ofp-www.jcahpc.jp/}}

\bibitem{KNLtextbook}
A.S. J.~Jeffers, J.~Reinders, \emph{Intel Xeon Phi Processor High Performance
  Programming Knights Landing Edition} (Elsevier, 2016)

\bibitem{thiscontrib_intel}
H.~Servat, \emph{{Getting the most out of Intel(R) Xeon Phi(tm) processor }},
  in \emph{Proceedings, \href{http://inspirehep.net/record/1425631}{35th
  International Symposium on Lattice Field Theory (Lattice2017)}: Granada,
  Spain}, to appear in EPJ Web Conf.

\bibitem{Motoki:2016rii}
S.~Motoki et~al., PoS \textbf{LATTICE2015}, 040 (2016)

\bibitem{bridge_alt}
{H. Matsufuru, S.Aoki, T.Aoyama ,K.Kanaya, S.Motoki, Y.Namekawa, H. Nemura,
  Y.Taniguchi, S.Ueda, and N. Ukita} (Bridge++ Project), Procedia Computer
  Scienc \textbf{51}, 1313 (2015)

\bibitem{Boyle:2016lbp}
P.A. Boyle, G.~Cossu, A.~Yamaguchi, A.~Portelli, PoS \textbf{LATTICE2015}, 023
  (2016)

\bibitem{grid}
\urlstyle{tt}\url{{https://github.com/paboyle/Grid}}

\bibitem{cander2017}
I.~Kanamori, H.~Matsufuru (2017), to appear in \emph{Proceedings, 5th
  International Workshop on Legacy HPC Application Migration (LHAM'17) in The
  fifth International Symposium on Computing and Networking (CANDAR'17):
  Aomori, Japan}

\end{thebibliography}

\end{document}